\documentclass[twocolumn,
 preprintnumbers,
 nofootinbib,
 amsmath,amssymb,
 prl,
]{revtex4-2}

\usepackage{graphicx}
\usepackage{dcolumn}
\usepackage{bm}
\usepackage[colorlinks,citecolor=blue,pdfencoding=auto]{hyperref}
\usepackage{xcolor}
\usepackage{soul}
\usepackage{tikz-feynman}
\usepackage{slashed}
\usepackage{multirow}
\usepackage{braket}
\usepackage{subcaption}
\usepackage{comment}
\def\tt{\tau^-\tau^+}

\newcommand{\tr}{\operatorname{tr}}
\usepackage[normalem]{ulem}

\begin{document}
\title{Decoherence and More Coherence in the Radiative Decay of the $Z$ Boson}

\author{Kun Cheng}
\email{kun.cheng@pitt.edu}
\affiliation{PITT PACC, Department of Physics and Astronomy,\\ University of Pittsburgh, 3941 O’Hara St., Pittsburgh, PA 15260, USA}

\author{Tao Han}
\email{than@pitt.edu}
\affiliation{PITT PACC, Department of Physics and Astronomy,\\ University of Pittsburgh, 3941 O’Hara St., Pittsburgh, PA 15260, USA}

\author{Harman Singh}
\email{HPS25@pitt.edu}
\affiliation{PITT PACC, Department of Physics and Astronomy,\\ University of Pittsburgh, 3941 O’Hara St., Pittsburgh, PA 15260, USA}

\author{Youle Su}
\email{YOS76@pitt.edu}
\affiliation{PITT PACC, Department of Physics and Astronomy,\\ University of Pittsburgh, 3941 O’Hara St., Pittsburgh, PA 15260, USA}

\begin{abstract}
Final state radiation in collider processes can be interpreted as interactions with unobserved degrees of freedom and is often discussed within the context of decoherence of an entangled state. 
We consider the radiative decay of the $Z$ boson as a representative example and perform a detailed analytical study of the $\tau^-\tau^+\gamma$ spin state, as determined by the chiral interactions of the Standard Model, over the complete three-body phase space.
We explore various quantum information observables to quantify how the $\tau^-\tau^+$ spin state changes with the photon radiation.
We find some striking features, for example that the emitted photon could either lead to decoherence or monotonic enhancement of entanglement for the fermion pair. 
\end{abstract}

\preprint{PITT-PACC-2609}

\maketitle

\noindent
\emph{Introduction ---}
Quantum entanglement is one of the most distinct phenomena in quantum mechanics.
In recent years, it has been shown that elementary particles in high energy collider experiments can make compelling systems for studying quantum information~\cite{Afik:2020onf,Fabbrichesi:2021npl,Severi:2021cnj,Afik:2022kwm,Han:2023fci,Aguilar_Saavedra_2022,Dong_2024,Cheng:2023qmz,Cheng:2024btk,Altakach:2022ywa,Ehataht:2023zzt,Ma:2023yvd,Fabbrichesi:2024wcd,Han:2025ewp,Zhang:2025mmm,Ai:2025wnt,Barr:2021zcp,Barr:2022wyq,Ashby-Pickering:2022umy,Aguilar-Saavedra:2022wam,Fabbrichesi:2023cev,Fabbri:2023ncz,Bi:2023uop,Morales:2023gow,Bernal:2024xhm,Cheng:2024rxi,Grossi:2024jae,Goncalves:2025qem,Goncalves:2025xer,Aguilar-Saavedra:2022mpg,Du:2024sly,Afik:2025grr,Chen:2013epa,Pei:2025ito,Fucilla:2025kit,Cheng:2025cuv,Lin:2025eci,Cheng:2025zcf,Guo:2026yhz,Fang:2026ddi,Qi:2025onf,Cheng:2025zaw,Hatta:2025obw,Fucilla:2025kit,Zhang:2026nwm,Goncalves:2026njf,Cheng:2026ktp}.
The observation of quantum entanglement of top quark pairs produced at the Large Hadron Collider constitutes the highest energy measurement of entanglement ever achieved~\cite{ATLAS:2023fsd,CMSCollaboration_2024}, highlighting the greater potential of high energy colliders to investigate other quantum phenomena through various unique processes.

Decoherence is a process in which a quantum state loses its quantum nature like entanglement. It is of fundamental importance to comprehend and control quantum decoherence in quantum information science and quantum computation. It is valuable to understand the decoherence phenomena in high energy physics processes. 
Final state radiation (FSR) of photons or gluons is often associated with quantum decoherence, since it can be interpreted as interactions with unobserved environmental degrees of freedom. 
It is generally expected that the emission of a photon or gluon alters the original quantum state of the particle pair, leading to quantum decoherence and reducing the entanglement after tracing over the spin degrees of freedom of the emitted quanta. 
It was argued that the unresolved radiation preserves the spin coherence due to the soft theorem \cite{Carney:2017jut}. 
There has been a growing interest in how additional radiation affects bipartite entanglement using the decoherence framework~\cite{Aoude:2025ovu,Gu:2025ijz,Aoude:2026eeg}.

In this work, we quantitatively examine the impact of hard radiation on entangled fermion-antifermion pairs. As an example, we present a detailed analytical study of the final spin states of the radiative decay $Z\to \tau^-\tau^+\gamma$ with respect to the complete three-body kinematics.
Rather than examine the three-body entanglement~\cite{Morales:2024jhj,Sakurai:2023nsc,Goncalves:2026nnx}, we investigate the bipartite entanglement between $\tau^-$ and $\tau^+$ by tracing out the emitted photon. Moreover, for each $Z\to \tau^-\tau^+$ decay angle, we compare the spin state of $\tau^-\tau^+$ with or without the radiated photon with varying photon energies.
In stark contrast to the common expectation, we find that hard photon radiation does not necessarily decrease the entanglement between $\tau^-$ and $\tau^+$. Instead, for some specific radiation angles, the entanglement between the $\tau$ pair even increases monotonically with the radiation energy as a result of the symmetry of dynamics and kinematics. This observation may provide a significant insight into understanding the phenomenon of decoherence in high energy physics processes.

\vspace{5pt}
\noindent
\emph{Entanglement of $\tt$ from Z Decay ---}
To set the stage, we start from an on-shell $Z$-boson decay to a pair of $\tau$ leptons. The tree-level helicity amplitude of $Z_{\sigma}\to \tau^-_a\tau^+_{\bar a}$ is expressed as
\begin{equation}
\label{eq:amplitude2body}
    \mathcal{M}_{ a \bar{a}}^{\sigma} = 
    { e
    \over s_W c_W}\ \epsilon^\sigma_{Z,\mu}  \bar{u}_a 
    \gamma^\mu \left( g_L^{\tau} P_L + g_R^{\tau} P_R\right) v_{\bar{a}}.
\end{equation}
Here, $\epsilon_Z^\sigma$ is the polarization vector that can be replaced with any vector current from which the $Z$ boson is produced. The spin index of the $Z$ boson is $\sigma=0,\pm1$, reducing to $\pm 1$ if the $Z$ boson is produced by a pair of massless fermions. The spins of $\tau^-$ and $\tau^+$ are denoted by the indices $a$ and $\bar a$.
The left and right chiral projection operators are $P_{L/R} = \frac{1}{2}(\mathbb{I} \mp \gamma_5)$, with the corresponding neutral current couplings $g_{L}^{e/\tau}=-\frac{1}{2}+s_W^2$ and $g_R^{e/\tau}=s_W^2$ with $s_W$/$c_W$ the sine/cosine of the weak mixing angle. 


When the on-shell $Z$-boson is produced by a pair of massless fermions, its polarization vector is given by the incoming current
\begin{equation}\label{eq:epsZfromee}
\bar{v}_{L/R} \gamma_\mu u_{R/L} = m_Z \epsilon_{Z,\mu}^{\pm}~ .   
\end{equation}
The spin direction $\vec S$ of the $Z$ boson in its rest frame is along the momentum of the initial fermion beam. For example, the complete $2\to 2$ amplitude of $e^-_Le^+_R/e^-_Re^+_L \to Z\to \tau^-_a\tau^+_{\bar a}$ is related to the decay amplitudes of $Z$ as
\begin{equation}\label{eq:ampEEampZ}
    \mathcal{M}^{RL/LR}_{ a\bar a} =\frac{g_{R/L}^{e}e}{s_W c_W}  \cdot\Delta \cdot  \mathcal{M}^{\pm}_{ a\bar a},
\end{equation}
where $\Delta= m_Z/(s-m_Z^2+im_Z\Gamma_Z)\approx -i/\Gamma_Z$. The process can be specified by two kinematical variables  $\beta_\tau = \sqrt{1-4m_\tau^2/m_Z^2}$ and $\theta$, the $\tau^-$ scattering angle with respect to the $e^-$ beam. 
This factorized form is general for any on-shell $Z$ production and applicable to $pp\to Z\to\tt(\gamma)$ as well at the LHC.

Consider the $\tau^-\tau^+$ pair as a bipartite two-qubit system, the spin density matrix can be written as 
\begin{equation}\label{eq:RhoDecomp}
\begin{aligned}
    \rho_{\tau\tau} & = \frac{I_4 + B_i^- \sigma_i \otimes I_2 +  B_j^+ I_2 \otimes\sigma_j + C_{ij}\sigma_i \otimes \sigma_j }{4}.\\
\end{aligned}
\end{equation}
Quantum tomography determines the 15 parameters of the polarizations $B^\pm_j$ and the correlation coefficients $C_{ij}$. The two-body density matrix of $\tau^-\tau^+$ is given by
\begin{equation}
    \rho_{a\bar a,b\bar b}=\frac{\sum \left(\mathcal{M}_{ b \bar{b}}^{\sigma}\right)^*\mathcal{M}_{ a \bar{a}}^{\sigma}}{\sum \left|\mathcal{M}_{ a \bar{a}}^{\sigma}\right|^2} \ket{a\bar{a}}\bra{b\bar{b}}
\end{equation}
where the initial beam polarizations are summed over, resulting in a mixed state.
The mixture is measured by the purity $\xi \equiv \tr(\rho^2)$ where $\xi=1$ corresponds to a pure state. The spin state of $\tau^-\tau^+$ from $Z$ decay occupies only two dimensions of the four-dimensional Hilbert space, namely $\ket{\uparrow\uparrow}$ and $\ket{\downarrow\downarrow}$, since the helicity flipping terms are suppressed by the small $\tau$ lepton mass. Therefore, purity is bounded by $1/2 \lesssim \xi \leq 1$. 

\begin{figure}[tb] 
    \centering
    \includegraphics[]{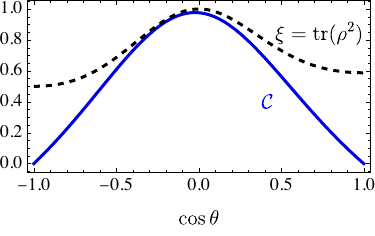}
    \caption{Purity $\xi$ and concurrence $\mathcal{C}$ of the spin state of $\tt$ from the $Z$ boson decay.}
    \label{fig:concurrenceWITHOUTradiation}
\end{figure}

We use the concurrence of the bipartite system as a measure of entanglement~\cite{Wootters:1997id}, defined as $\mathcal{C}=\max\!\left\{0,\lambda_1-\lambda_2-\lambda_3-\lambda_4\right\}$ where $\lambda_i$ are the eigenvalues in the descending order of the auxiliary matrix $R=\sqrt{\sqrt{\rho} \tilde{\rho} \sqrt{\rho}}$ where $\tilde{\rho}=(\sigma_2 \otimes \sigma_2) \rho^* (\sigma_2 \otimes \sigma_2)$. 
The concurrence for the massless final state in terms of $\theta$ is approximately given by~\cite{Cheng:2024rxi,Guo:2026yhz} 
\begin{equation}
 \mathcal{C}\approx {\sin^2\theta \over {1+\cos^2\theta} }\ .
\end{equation} 
We present $\xi$ and $\mathcal{C}$ in Fig.~\ref{fig:concurrenceWITHOUTradiation}.
The chiral coupling of the $Z$ boson to the charged leptons exhibits the approximate relation $|g_R|\approx |g_L|$. The amplitudes of producing $\ket{\uparrow\uparrow}$ and $\ket{\downarrow\downarrow}$ states are approximately equal in the central scattering limit, resulting in a Bell state with maximal concurrence and purity, while both concurrence and purity reach their minimum in the forward/backward region where the spin state is a classical mixture of $\ket{\uparrow\uparrow}$ and $\ket{\downarrow\downarrow}$ with about the same probability.
In the following, we explore how hard radiations modify the entanglement and purity, in connection with decoherence.

\begin{figure}
  \centering
    \begin{tikzpicture}[baseline=(v1.base)]
      \begin{feynman}
        \vertex (v1) at (0, 0);
        \vertex (e1) at (-1.,  1.0) {\(e^{-}/q\)};
        \vertex (e2) at (-1., -1.0) {\(e^{+}/\bar q\)};
        \vertex (v2) at ( 0.8,  0.0);
        \vertex (tp) at ( 2.,  1.2) {\(\tau^{+}\)};
        \vertex (v3) at ( 1.25, -0.4);
        \vertex (tm) at ( 2.2, -1.2) {\(\tau^{-}\)};
        \vertex (gam) at (2.2,  0.2) {\(\gamma\)};

        \diagram* {
          (e1) -- [fermion] (v1),
          (v1) -- [fermion] (e2),
          (v1) -- [boson, edge label=\(Z\)](v2),
          (tp) -- [fermion, edge label'=\(p_+\)]  (v2),
          (v2) -- [fermion] (v3),
          (v3) -- [fermion, edge label'=\(p_-\)]  (tm),
          (v3) -- [boson,   edge label=\(k\)]    (gam),
        };
      \end{feynman}
    \end{tikzpicture}
  +
    \begin{tikzpicture}[baseline=(v1.base)]
      \begin{feynman}
        \vertex (v1)  at (0,    0.0);
        \vertex (e1)  at (-1,  1.0) {\(e^{-}/q\)};
        \vertex (e2)  at (-1, -1.0) {\(e^{+}/\bar q\)};
        \vertex (v2)  at ( 0.8,  0.0);
        \vertex (tm)  at ( 2, -1.2) {\(\tau^{-}\)};
        \vertex (v3)  at ( 1.24,  0.4);
        \vertex (tp)  at ( 2.2,  1.2) {\(\tau^{+}\)};
        \vertex (gam) at ( 2.2, -0.2) {\(\gamma\)};

        \diagram* {
          (e1) -- [fermion] (v1),
          (v1) -- [fermion] (e2),
          (v1) -- [boson, edge label=\(Z\)] (v2),
          (v2) -- [fermion, edge label'=\(p_-\)] (tm),
          (v2) -- [anti fermion] (v3),
          (tp) -- [fermion, edge label'=\(p_+\)] (v3),
          (v3) -- [boson, edge label'=\(k\)] (gam),
        };
      \end{feynman}
    \end{tikzpicture}

  \caption{Leading-order Feynman diagrams for $e^+e^-, q \bar q\to Z\to\tau^+\tau^-\gamma$}
  \label{fig:feynman-diagrams}
\end{figure}
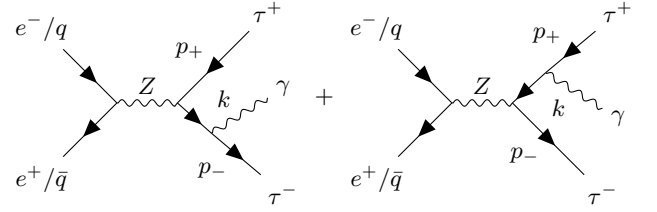

\begin{figure}
    \centering
    \includegraphics[width=\linewidth]{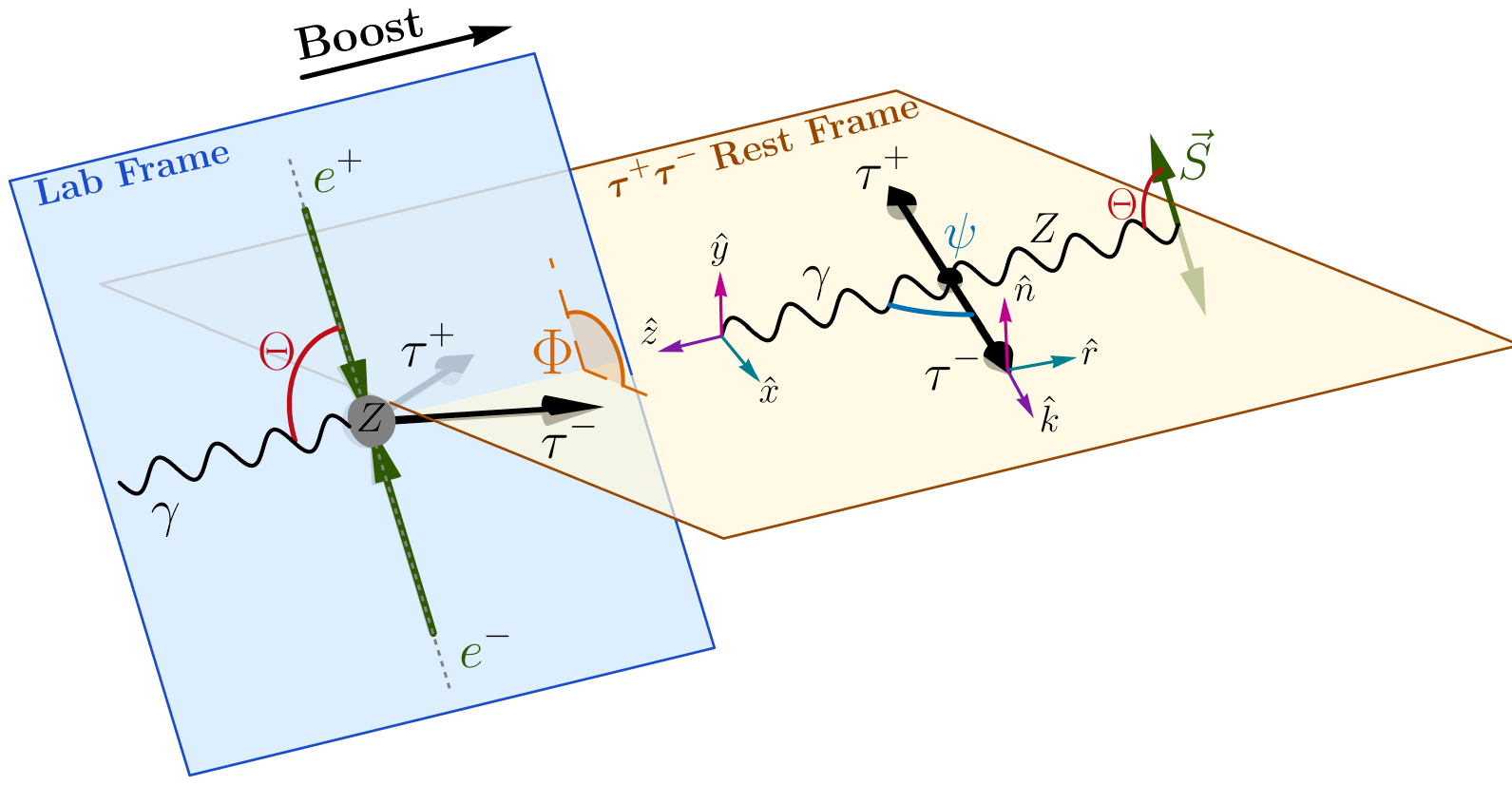}
    \caption{Illustration for the $2\to 3$ kinematics in the lab frame (the $Z$ rest frame) and $\tt$ c.~m.~frame. $\vec S$ is the spin vector of $Z$ in its rest frame.
    }
    \label{fig:kinematics}
\end{figure}
\vspace{5pt}
\noindent
\emph{Three Body Kinematics ---}
Consider a $Z$ boson decaying to a pair of $\tau$ leptons plus a photon as depicted in Fig.~\ref{fig:feynman-diagrams}, where the radiated photon can be hard and with a finite opening angle. 
It is convenient to divide the full tree-body phase space of $\tt \gamma$ into two factors: the two-body phase space of the photon and the $\tt$ system, and then the two-body phase space of $\tt$ in their center-of-mass (c.~m.)~frame. The decay kinematics are shown in Fig.~\ref{fig:kinematics} in both the 
$Z$ rest frame and the $\tt$ c.~m.~frame. 
First, the two-body phase space of the photon and $\tt$ system is described by the scattering angle $\Theta$ between the photon and the incoming momentum of the electron, and the $\tau$ pair invariant mass $m_{\tau\tau}$. This two-body phase space defines the blue scattering plane in Fig.~\ref{fig:kinematics}. The photon energy in the $Z$ rest frame is related to $m_{\tau\tau}$ as 
\begin{equation}
    E_\gamma=(1-x^2){m_Z\over 2},\quad x={m_{\tau\tau} \over m_Z}.
\end{equation}
Second, the two-body phase space of $\tt$ is described by a polar angle $\psi$ of the photon with respect to the $\tau^-$ direction in the $\tt$ c.~m.~frame and an azimuthal angle $\Phi$ between the scattering planes of $e^-\gamma\ (\Phi=0)$ and $\tau^-\gamma$. These two coordinate frames coincide in the soft radiation limit $E_\gamma \to 0$, and the three-body kinematics restores to the two-body phase space of $\tau$ pair from $Z\to \tau^-\tau^+$, with the $\tau^-$ scattering angle $\theta$ given by 
\begin{equation}\label{eq:222theta}
   \cos\theta = \cos\psi  \cos\Theta 
+\sin\psi  \sin\Theta  \cos\Phi .
\end{equation}

\noindent
\emph{Amplitudes --- }
The tree-level helicity amplitude of $Z\to \tt \gamma$ is given by 
\begin{widetext}
\begin{equation}
\label{eq:amplitudes}
    \mathcal{M}_{\lambda a \bar{a}}^{\sigma} = 
    { e^2 \over s_W c_W}\ \epsilon^\sigma_{Z,\mu}  \bar{u}_a 
    \left( \slashed{\varepsilon} ^*_\lambda \frac{\slashed{p}_-+\slashed{k} + m_\tau}{\left(p_-+k \right)^2-m_\tau^2} \gamma^\mu \left( g_L^{\tau} P_L + g_R^{\tau} P_R\right) - \gamma^{\mu} \left(g_L^{\tau} P_L + g_R^{\tau} P_R\right)  \frac{\slashed{p}_++\slashed{k} - m_\tau}{\left(p_++k \right)^2-m_\tau^2} \slashed{\varepsilon} ^*_\lambda   \right) v_{\bar{a}}.
\end{equation}
\end{widetext}
where $p_\mp$ are the four-momenta of the outgoing $\tau^{\mp}$. For $e^-e^+$ annihilation, the polarization vector of the $Z$ boson is given by Eq.~\eqref{eq:epsZfromee} and the complete amplitude is related to the decay amplitude as in Eq.~\eqref{eq:ampEEampZ}.

As shown in Fig.~\ref{fig:kinematics}, the angles $\Theta$ and $\Phi$ are also the spin direction of the $Z$ boson with respect to the photon in the $\tau^-\tau^+$ c.~m.~frame. Therefore, all the $\Theta$ and $\Phi$ dependence of the amplitudes are in the form of Wigner-$D$ functions. For $e^-_R e^+_L$ scattering, the amplitudes can be written in the massless $\tau$ limit as
\begin{equation}
    \mathcal{M} =\frac{4\sqrt{2}e^3f_j^R}{m_Z(1-x^2)}\left(\sum_{i=-1}^1 F_i D_{i,1}^1(\Phi,\Theta)\right)
\end{equation}
where $f_{j}^i$ are the chiral factors, 
\begin{equation}
\label{eq:fij}
    f_{j}^i= \frac{g_i^e g_j^\tau}{2c_W^2 s_W^2} \frac{m_Z}{i\Gamma_Z}\quad i,j=L,R,
\end{equation}
and $D_{i,1}^1(\Phi,\Theta)$ are the Wigner-$D$ functions given in the Appendix.
The coefficients $F_{\pm}$ are proportional to the decay amplitude of the $Z$ boson whose spin can be along or opposite to the photon momentum direction, while $F_{0}$ corresponds to the amplitude of the $Z$ boson whose polarization is longitudinal to the photon momentum direction. 

\begin{table}[tb]
    \centering
    \tabcolsep=0.15cm 
    \renewcommand\arraystretch{1.2} 
    \begin{tabular}{|c|c|c|c|c|}
        \hline
        $\tau^-_a \tau^+_{\bar a}\gamma_\lambda$ &$f_{j}^R$& $F_1$ & $F_0$ & $F_{-1}$ \\
        \hline
        $\uparrow\uparrow+$ &$f_R^R$& $-\cot{(\psi/2)}$ & $-\sqrt{2}x$ &$-x^2\tan{(\psi/2)}$\\
        \hline
        $\downarrow\downarrow+$ &$f_L^R$& $-\tan{(\psi/2)}$ & $\sqrt{2}x$ & $-x^2\cot{(\psi/2)}$ \\
        \hline
        $\uparrow\uparrow-$ &$f_R^R$& $x^2\cot{(\psi/2)}$ & $\sqrt{2}x$ & $\tan{(\psi/2)}$ \\
        \hline
        $\downarrow\downarrow-$ &$f_L^R$& $x^2\tan{(\psi/2)}$ & $-\sqrt{2}x$ & $\cot{(\psi/2)}$ \\
        \hline
    \end{tabular}
        \caption{The coefficients $F_i$ for massless fermions with the $e_R^- e_L^+$ initial state.}
    \label{table:amp}
\end{table}

The helicity amplitudes of $e^-_Re^+_L\to Z\to \tau^-_a\tau^+_{\bar a}\gamma_{\lambda}$ with massless $\tau$ leptons are listed in Table~\ref{table:amp} and the results for massive $\tau$ leptons are listed in the Appendix. The amplitude of $e_L^- e_R^+$ scattering is obtained by flipping the spin direction of $\epsilon_Z$, {\it i.e.}, $\Theta\to \pi-\Theta$ and $\Phi\to \pi+\Phi$, and replacing the initial state chiral coupling $f_{j}^R$ with $f_{j}^L$. 
The spins of the final state $\tau^-$ and $\tau^+$ are defined in the $\tt$ c.~m.~frame, both quantized along the $\tau^-$ moving direction. This corresponds to the ``helicity basis" $(r,n,k)$ in Fig.~\ref{fig:kinematics}. Alternatively, the ``photon momentum basis" $(x,y,z)$ is a convenient choice near threshold $m_{\tau\tau}\approx  2m_\tau$. In this regime, the photon energy is large ($x\to 2m_\tau/m_Z \ll 1$), and the only surviving amplitudes are those when the $Z$ spin aligns with the photon's spin.

Similarly to three-jet events from gluon radiation off quarks \cite{Ellis:1976uc,TASSO:1979zyf}, the radiated photon tends to be coplanar with the $\tau$ leptons $\Phi \to 0$ or $\pi$, as discussed in detail in the Appendix. For the non-coplanar subdominant region, we can see from Table \ref{table:amp} that they will exhibit the same qualitative features as the coplanar case. As an example, consider $\Phi=\pi/2$. In the collinear limit, the $\Phi$-dependence of the amplitudes appears only as an overall phase factor and therefore does not modify the spin state. In contrast, for the interesting configuration $\psi\to \pi/2$, the angle $\theta$ is constrained as seen in Eq.~\eqref{eq:222theta}.
Therefore, we will focus on the dominant coplanar configuration $\Phi = 0, \pi$, where most events are populated and the available phase space is largest. 

\vspace{5pt}
\noindent
\emph{Entanglement Between $\tau^+$ and $\tau^-$ ---}
To characterize the decoherence effects, we start by considering the bipartite entanglement and purity of the $\tau^-\tau^+$ state, and focus on the difference between the states with and without photon radiation. The three-body kinematics is thus specified by $\Theta(\gamma e^-)$, $\psi(\gamma \tau^-)$ and $E_\gamma$ (or $x=m_{\tau\tau}/m_Z$). To gain an intuitive comparison with the kinematics of $Z\to \tt$ without photon radiation,  in the following we present the spin state as a function of $\theta(\tau^- e^-)$ in Eq.~\eqref{eq:222theta} instead of $\Theta(\gamma e^-)$. 

\begin{figure}
    \centering
    \includegraphics[width=\columnwidth]{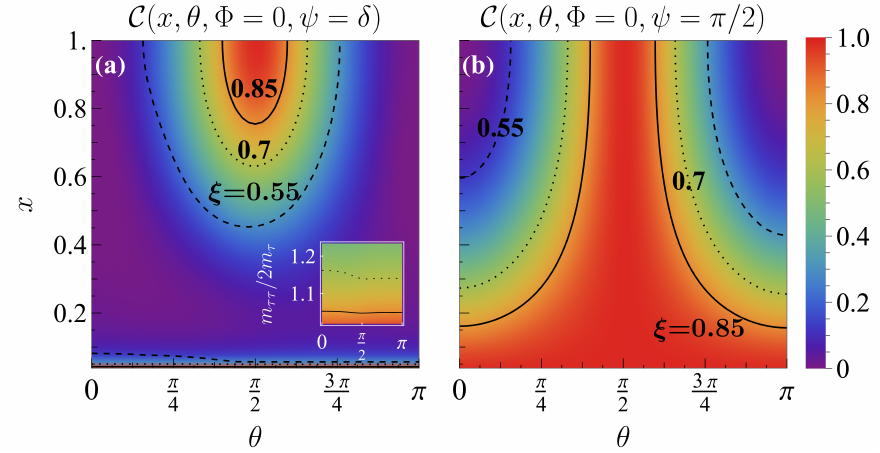}
    \caption{ Concurrence (color density) and purity (contour lines) of massive $\tau^-\tau^+$ in the plane $(\theta,x)$. Photon radiation angle is (a) collinear with (b) perpendicular to the $\tau$ leptons in $\tt$ c.~m.~frame. The zoom-in subplot in (a) shows the threshold behavior when $m_{\tau\tau}\approx 2m_\tau$.}
    \label{fig:coplanar_conc_massive_EW}
\end{figure}

As seen in Table \ref{table:amp}, in the collinear radiation limit $(\psi\to0,\pi)$, the massless amplitudes diverge. We thus adopt an angular cutoff $\delta$ on $\psi$ as 
\begin{equation}
    \delta=0.1, 
\end{equation}
which regularizes the collinear photon radiation and also implies a separation cut for observing the photon-lepton events. 
Perpendicular radiation $(\psi\to \pi/2)$ is an interesting opposite limit that has not been appreciated. We demonstrate these two characteristic cases in the following. 
In Fig.~\ref{fig:coplanar_conc_massive_EW}, we show the bipartite concurrence (color density) and purity ($\xi$ contour lines) of $\tau^-\tau^+$ in the plane of $(\theta,x=m_{\tau\tau}/m_Z)$, with the $Z$ boson produced from unpolarized $e^-e^+$ annihilation and photon polarization traced over. 
As stated in the Weinberg soft theorem, when the photon is soft as $x=m_{\tau\tau}/m_Z\to 1$, the entanglement between $\tau^-\tau^+$ restores to $Z\to \tau^-\tau^+$ in Fig.~\ref{fig:concurrenceWITHOUTradiation} regardless of the direction of the radiated photon.

Starting from collinear radiation, Figure \ref{fig:coplanar_conc_massive_EW}(a) shows a clear comparison of the concurrence of the $\tau^-\tau^+$ system with/without photon radiation with decreasing $x$ (increasing radiated photon energy). 
We see that the entanglement of $\tau^-\tau^+$ decreases with the radiation energy, leading to a qualitatively expected decoherence effect similar to the results in Ref.~\cite{Aoude:2025ovu,Gu:2025ijz,Aoude:2026eeg}. This is transparent from Table~\ref{table:amp} where we see the coherence superposition between $\ket{\uparrow\uparrow}$ and $\ket{\downarrow\downarrow}$ decreases as $x^2$. When the photon energy reaches maximum, the entanglement increases again as shown in the zoom-in plot of Fig.~\ref{fig:coplanar_conc_massive_EW}(a). This corresponds to a special limit only for a massive fermion pair where the $\tau$ pair spin state is maximally entangled independent of the photon emission angle. The same feature of decoherence and re-coherence is observed in Ref.~\cite{Aoude:2026eeg} for the $t\bar t g$ channel. 
Comparing the color density with the contour lines, we also observe that the concurrence is strongly correlated with the purity. This indicates that the entanglement of the $\tau$ pair decreases only when their spin states associated with positively and negatively polarized photon emission differ. Therefore, tracing over the photon polarizations produces a mixed state, reducing the purity and entanglement simultaneously.

In contrast, if the angle of photon radiation is $\pi/2$ in the $\tt$ c.~m.~frame (equivalently $\tau^-$ and $\tau^+$ have the same energy in the $Z$ rest frame), the entanglement monotonically increases with the radiation energy, as shown in Fig.~\ref{fig:coplanar_conc_massive_EW}(b).
The absence of decoherence when $\psi=\pi/2$ and $\theta=\pi/2$ is a result of an approximate charge conjugation symmetry of the system. The $Z$ boson couples to charged leptons dominantly through an axial vector current, with nearly equal couplings to $\tau_L^-\tau_R^+$ and $\tau_R^-\tau_L^+$. Consequently, both the scattering kinematics and the interactions have a charge conjugation symmetry. The approximately same amplitudes of $\tau_L^-\tau_R^+$ and $\tau_R^-\tau_L^+$ lead to a maximally entangled state $(\ket{\uparrow\uparrow}-\ket{\downarrow\downarrow})/\sqrt{2}$.
In addition, Table \ref{table:amp} shows the amplitude for energetic photons is dominated by terms where the photon and $Z$ boson have the same spin direction. With the spins of both bosons perpendicular to the $\tau$ leptons in $\tau^-\tau^+$ c.~m.~frame, the system exhibits a charge conjugation symmetry regardless of $\theta$. Consequently, the entangled region is expanded to the forward/backward regions, as shown in Fig.~\ref{fig:coplanar_conc_massive_EW}(b) with decreasing $x$. 
In comparison, the kinematics in the collinear limit does not respect the charge conjugation symmetry so the spin configuration is no longer protected and the entanglement decreases with increasing photon energy.

Finally, both plots in Fig.~\ref{fig:coplanar_conc_massive_EW} show that, in the threshold limit $m_{\tau \tau} \to 2 m_\tau$  ($E_\gamma  \sim m_Z/2$), the $\tau$ pair is nearly maximally entangled regardless of any other kinematics. 
In this limit, the photon is sufficiently energetic to kick the $\tau$ lepton pair to be co-moving without any relative velocity. Therefore, the $\tau$ pair can be treated as a spin-1 system with zero orbital angular momentum. Conservation of angular momentum then requires that the photon have the same spin as the $Z$ boson, constraining the $\tau$ pair to be spin-0 along the photon direction. With almost pure axial current coupling, the spin state is always close to the maximumly entangled $\ket{1,0}$ state under the photon momentum basis $(x,y,z)$ in Fig.~\ref{fig:kinematics}, and the spin state does not depend on the scattering angle, as seen in the bottom band of Fig.~\ref{fig:coplanar_conc_massive_EW}(a). Unlike the $Z\to \tt$ process, where the state is entangled in the central scattering region but separable in the forward/backward region, the $\tau$ pair is almost maximally entangled at any scattering angle when the energy of the radiated photon is close to maximum.

\begin{figure}
    \centering    
       \includegraphics[scale=0.9]{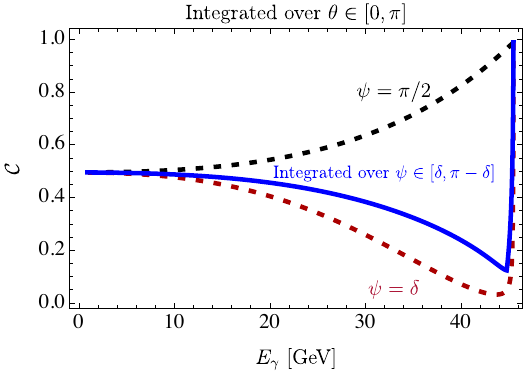}
    \caption{Concurrence integrated over $\theta\in[0,\pi]$ for $\psi=\delta$ (bottom-dashed curve), $\pi/2$ (upper dashed curve), and integrated $\psi$ (middle solid curve). }
    \label{fig:coplanar_conc_int2}
\end{figure}

To achieve a global picture, we show the concurrence of the angular average over $\theta$ and $\psi$ as a function of photon energy in Fig.~\ref{fig:coplanar_conc_int2}. 
The bottom-dashed curve is for the collinear photon configuration, which exhibits the characteristic of decoherence-recoherence to increase $E_\gamma$.  The upper-dashed curve is for the perpendicular photon configuration, which shows the monotonic increase of entanglement over $E_\gamma$. The solid in the middle curve shows the integrated result in $\psi$, which is close to the collinear case.
For the $\psi$ integral, the spin state is averaged in the helicity or photon momentum basis, providing a greater concurrence.

\vspace{5pt}
\noindent
\emph{Quantum Magic ---}
Quantum magic, or non-stabilizerness, quantifies the departure of a state from the set of stabilizer states generated by Clifford gates, which can be efficiently simulated on classical computers~\cite{Nielsen:2012yss}.
For two-qubit systems, a typical measure of magic is the second Stabilizer R\'{e}nyi entropy (SSRE) $\mathcal{M}_2$. 
The SSRE is given in terms of its quantum tomography by~\cite{Leone:2021rzd,White:2024nuc}
\begin{equation}
    \mathcal{M}_2 = -\log_2 \left(\frac{1+\sum_i (B_i^-)^4+\sum_j (B_j^+)^4 + \sum_{ij} (C_{ij})^4}{1+\sum_i (B_i^-)^2+\sum_j (B_j^+)^2 + \sum_{ij} (C_{ij})^2} \right).
\end{equation}

\begin{figure}
    \centering
        \includegraphics[width=\columnwidth]{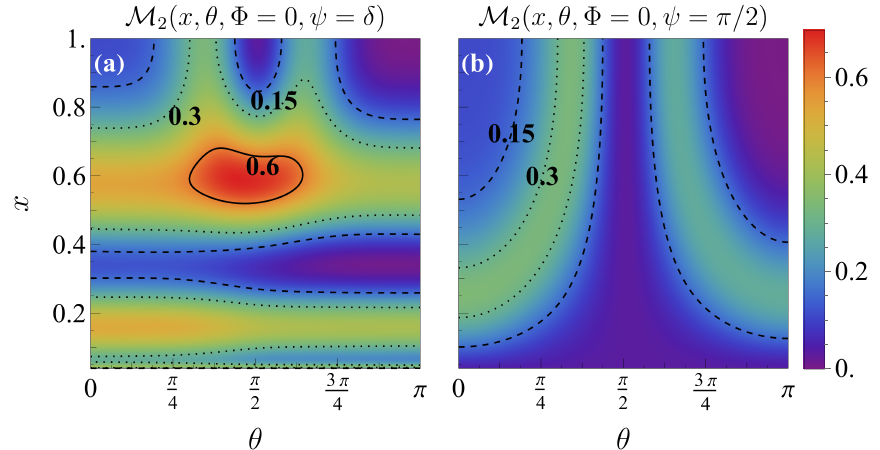}
    \caption{ Magic ($\mathcal{M}_2$, color density) of massive $\tau^+\tau^-$ as in Fig.~\ref{fig:coplanar_conc_massive_EW} }
    \label{fig:magic}
\end{figure}

In the two body process of $Z\to\tau^-\tau^+$, the state in the central scattering region is close to a maximally entangled Bell state $(\ket{\uparrow\uparrow}-\ket{\downarrow\downarrow})/\sqrt{2}$, while forward/backward scattering produces mixed and separable states $\rho=(\ket{\uparrow\uparrow}\bra{\uparrow\uparrow}+\ket{\downarrow\downarrow}\bra{\downarrow\downarrow})/2$. Both states are stabilizer states with zero magic.
It is therefore informative to explore magic to quantify the deviation from these states with a photon radiation.
Figure~\ref{fig:magic} shows the magic of the $\tau^-\tau^+$ system in the presence of photon radiation with the same kinematics as in Fig.~\ref{fig:coplanar_conc_massive_EW}. 
Starting from the limit of $x \to 1$ that restores the two body decay $Z\to \tau^-\tau^+$, we see that magic is close to zero in the central and forward/backward scattering region as discussed above. We also see a forward-backward asymmetry of the magic compared to concurrence. For backward scattering ($\theta = \pi$), the $\tau^-$ and $e^-$ leptons always have opposite helicities, so the amplitudes for the $\ket{\uparrow\uparrow}$ and $\ket{\downarrow\downarrow}$ states are equal ($g_Lg_R$), resulting in a stabilizer state. In forward scattering ($\theta =0$), the $\tau^-$ and $e^-$ leptons always have the same helicity, so the amplitudes for the $\ket{\uparrow\uparrow}$ and $\ket{\downarrow\downarrow}$ states slightly differ since $g_L^2 \sim g_R^2$, giving a close to stabilizer final state with nonzero magic. In both regions, the states are not entangled, as they are incoherent mixtures of $\ket{\uparrow\uparrow}$ and $\ket{\downarrow\downarrow}$.

When the photon is radiated close to the collinear limit  as shown in Fig.~\ref{fig:magic}(a), the magic increases with decreasing $x$. This implies that the spin state of $\tau^-\tau^+$ always changes after the photon radiation, therefore deviating from the stabilizer state in the central and forward/backward regions. Near $x\sim 0.6$, magic becomes sizable and reaches 0.6 in the central region. 
As the photon becomes more energetic, the $\tau$ mass becomes relevant and the helicity flipping terms become non-negligible when $x\delta \sim 2 m_\tau/m_Z$.  When the amplitudes of $\ket{\uparrow\downarrow/\downarrow\uparrow}$ become comparable to  $\ket{\uparrow\uparrow/\downarrow\downarrow}$, the spin states approach a separable stabilizer state, as shown in the purple band of small magic around $x \sim 0.35$. As the energy of the $\tau$ pair approaches threshold, the bipartite system becomes re-entangled, so the spin states move away from the intermediary stabilizer state towards another stabilizer state $\frac{1}{\sqrt{2}}\left( \ket{\uparrow \downarrow } + \ket{\downarrow \uparrow} \right)$.

In the case where the photon is radiated perpendicular to the $\tau$ pair as shown in Fig.~\ref{fig:magic}(b), the states produced in the central scattering region are close to $\frac{1}{\sqrt{2}}\left( \ket{\uparrow \uparrow } - \ket{\downarrow \downarrow} \right)$. As discussed in the previous section, this spin configuration is protected by symmetry and remains in the entangled stabilizer state regardless of the photon energy. In the forward and backward regions, the state transitions from separable stabilizer states to the same entangled stabilizer state as the central region when the photon energy increases, resulting in the magic increasing as the state transitions near $x\sim 0.2$ between the two stabilizer states.

\vspace{5pt}
\noindent
\emph{Summary ---}
Final state radiation in collider processes can be interpreted as interaction with unobserved degrees of freedom and is naturally related to decoherence. 
In the case of photon radiation, different photon polarizations play the role of environmental degrees of freedom.
It is generally expected that a more energetic photon indicates a stronger interaction with the environment, leading to stronger decoherence.

We considered the radiative decay of $Z$ bosons and present a detailed analytical study of the $\tau^-\tau^+\gamma$ spin state with respect to the complete three-body kinematics.
In contrast to the common expectation and with the soft collinear limit, we found that hard photon radiation does not necessarily decrease the entanglement between $\tau^-$ and $\tau^+$. Instead, in some phase space region, the entanglement between the $\tau$ pair even monotonically increases with increasing radiation energy due to the symmetry of the dynamics and kinematics. 

Firstly, when the photon is emitted close to collinear with $\tau$, this corresponds to the configuration in which $\tau^-$ and $\tau^+$ exhibit the largest kinetic energy difference in the $Z$ rest frame.
As shown in Fig.~\ref{fig:coplanar_conc_massive_EW}(a), the entanglement of $\tau^-\tau^+$ first decreases with increasing radiation energy with an expected decoherence effect, and then the re-coherence occurs when the photon energy approaches maximum near the threshold $m_{\tau\tau}\approx 2m_\tau$.
Secondly, a striking feature is that if the photon is radiated perpendicular to the $\tau$ leptons in their c.~m.~frame, where $\tau^-$ and $\tau^+$ have the same energy in the $Z$ rest frame, the entanglement generally increases with increasing radiation energy, as shown in Fig.~\ref{fig:coplanar_conc_massive_EW}(b). This is a result of an approximate charge conjugation symmetry, and is not captured by the collinear behavior as discussed in the literature \cite{Aoude:2025ovu,Gu:2025ijz}.
Third, the threshold region for pairs of $\tau$ leptons $(m_{\tau\tau}\to 2m_\tau)$ represents a special kinematical limit where the photon reaches its maximum energy. In this limit, the $\tau$ lepton pair is almost maximally entangled regardless of other kinematics, due to the conservation of angular momentum, as seen in the small zoomed-in panel of Fig.~\ref{fig:coplanar_conc_massive_EW}(a), as well as in Fig.~\ref{fig:coplanar_conc_massive_EW}(b). 

Figure \ref{fig:coplanar_conc_int2} summarizes our main findings, with the bottom curve and the middle-solid curve showing the decoherence from collinear photon radiation to re-coherence near the threshold, and the upper curve showing the monotonic increase in entanglement from the perpendicular photon radiation. These features may change if there are new chiral interactions beyond the Standard Model.
Figure \ref{fig:magic} shows the rich structure of magic that reflects the stabilizerness of both the maximally entangled states and the separable states. 

As a final remark, although we took the annihilation of $e^-e^+$ as an example, our formalism and conclusions also hold for $Z$ bosons produced at the LHC, as well as for the other decay channels $Z\to e^-e^+\gamma$ and $Z\to \mu^-\mu^+\gamma$, subject to mass effects.

\vspace{10pt}
\noindent
\emph{Acknowledgments ---}
This work was supported in part by the US Department of Energy under grant N.~DE-SC0007914, and in part by Pitt PACC.

\appendix
\begin{widetext}
\section{End Matter: Helicity Amplitudes for Massive $f\bar f \gamma$}
In this appendix we list the most general helicity amplitudes for massive final states for the process $e^-e^+\rightarrow\gamma^*,Z^*\rightarrow f\bar{f}\gamma$. The kinematical variables are defined in Fig.~\ref{fig:kinematics}. 

For the $e_R^- e_L^+$ initial states, the $\Theta$ and $\Phi$ dependence of amplitudes are written as, 
\begin{equation}
    \mathcal{M}=\frac{\sqrt{2}e^3 P}{\sqrt{s}(1-x^2)(1-\beta^2\cos^2{\psi})}\left(F_1 D^1_{1,1}(\Phi,\Theta)+F_0 D^1_{0,1}(\Phi,\Theta)+F_{-1} D^1_{-1,1}(\Phi,\Theta)\right).
\end{equation}
The factor $P$ and $F_i$ are listed in  Table~\ref{table:amp_massive}, and  $D_{i,1}^1(\Phi,\Theta)$ are the Wigner-$D$ functions given by
\begin{equation}
        D_{\pm1,1}^1(\Phi,\Theta)=e^{\mp i\Phi}\frac{1\pm\cos{\Theta}}{2},\ 
D_{0,1}^1(\Phi,\Theta)=\frac{\sin{\Theta}}{\sqrt 2}.\quad  
\end{equation}
The chiral coupling functions in the table are
$k_R=\beta\left(f_V^R+\beta f_A^R\right)$, $k_L=\beta\left(f_V^R-\beta f_A^R\right)$,
$\kappa=f_A^R (1-x^2)(1-\beta^2)$ and $\kappa_{V/A\pm}=f_{V/A}^R(1-x^2)(1\pm\beta)$. The current and axial factors $f_{V/A}^i$ defined from the chiral factors $f_{V/A}^i=f_R^i\pm f_L^i$, with
\begin{equation}
\label{eq:fafv}
    f_{j}^i= \frac{1}{2}Q_e Q_f +  \frac{g_i^e g_j^f}{2c_W^2 s_W^2} \frac{s}{s-m_Z^2+is\Gamma_Z/m_Z},\quad i,j=L,R.
\end{equation}
where $\Gamma_Z$ is the $Z$ boson width at $Z$-pole and the fermion chiral couplings are
\begin{equation}
    g_L^f=I^3_f-Q_f s_W^2\quad  {\rm and}\quad g_R^f= -Q_f s_W^2.
    \label{eq:chiral_couplings}
\end{equation} 
$Q_f$ is the electric charge of the fermion $f$, and $I_{f}^3$ is the third component of the weak isospin.

For charged leptons, $Q_{\ell}=-1$ and $I_{f}^3=-1/2$ such that $g_L\approx - g_R$. 
At the $Z$ pole, the QED term $Q_e Q_f/2$ in Eq.~\eqref{eq:fafv} can be ignored for charged leptons and return to the chiral factors for $Z$ decay amplitude in Eq.~\eqref{eq:fij}. 

From Table~\ref{table:amp_massive}, we can immediately see the helicity flipped terms are suppressed by the final state mass since $P$ for the $\ket{\uparrow\downarrow}$ and $\ket{\downarrow\uparrow}$ sates is $\sqrt{1-\beta^2}\propto {m_f^2}/m_Z^2$. 
For the $LR$ initial states, the amplitude can be given by the substitution $\Theta\to \pi-\Theta$ and $\Phi\to \pi+\Phi$, and replacing the initial state chiral coupling $f_{j}^R$ with $f_{j}^L$. 

\begin{table}[htb]
    \caption{Helicity amplitude factors for the $RL$ initial states. }
    \label{table:amp_massive}
    \centering
    \tabcolsep=0.15cm 
    \renewcommand\arraystretch{1.2} 
    \begin{tabular}{|c|c|c|c|c|}
        \hline
        $f_a \bar f_{\bar a}\gamma_\lambda$ & $P$ & $F_1$ & $F_0$ & $F_{-1}$ \\
        \hline
        $\uparrow\uparrow+$ & $4\sin\left(\frac{\psi}{2}\right)$ & $- \left( \kappa + k_R \cos^2\left(\frac{\psi}{2}\right) \right)\cos\left(\frac{\psi}{2}\right)$ & $-\frac{x}{\sqrt{2}}  \left( \kappa + 2 k_R \cos^2\left(\frac{\psi}{2}\right) \right) \sin\left(\frac{\psi}{2}\right)$ &$-x^2k_R\sin^2\left(\frac{\psi}{2}\right)\cos\left(\frac{\psi}{2}\right)$\\
        \hline
        $\uparrow\downarrow+$ &$\sqrt{1-\beta^2}$& $2(\kappa_{A-}\cos{\psi}-\kappa_{V-}-f_V^R\beta\sin^2{\psi})$ & $\sqrt{2}x(\kappa_{A-}+2f_V^R \beta\cos{\psi})\sin{\psi}$ &$2f_V^R x^2\beta\sin^2{\psi}$\\
        \hline
        $\downarrow\uparrow+$ &$\sqrt{1-\beta^2}$& $2(\kappa_{A+}\cos{\psi}+\kappa_{V+}-f_V^R\beta\sin^2{\psi})$ & $\sqrt{2}x(\kappa_{A+}+2f_V^R \beta\cos{\psi})\sin{\psi}$ &$2f_V^Rx^2\beta\sin^2{\psi}$\\
        \hline
        $\downarrow\downarrow+$ & $4\cos{\left(\frac{\psi}{2}\right)}$ & $- \left(\kappa + k_L \sin^2\left(\frac{\psi}{2}\right) \right)\sin\left(\frac{\psi}{2}\right)$ & $-\frac{x}{\sqrt{2}} \left( \kappa - 2 k_L \sin^2\left(\frac{\psi}{2}\right) \right)\cos\left(\frac{\psi}{2}\right)$ & $-x^2 k_L \cos^2\left(\frac{\psi}{2}\right)\sin\left(\frac{\psi}{2}\right)$ \\
        \hline
        $\uparrow\uparrow-$&$4\cos{\left(\frac{\psi}{2}\right)}$ &$x^2k_R \cos^2\left(\frac{\psi}{2}\right)\sin\left(\frac{\psi}{2}\right)$ & $-\frac{x}{\sqrt{2}}\left( \kappa - 2 k_R \sin^2\left(\frac{\psi}{2}\right) \right)\cos\left(\frac{\psi}{2}\right)$ & $- \left( \kappa - k_R \sin^2\left(\frac{\psi}{2}\right) \right)\sin\left(\frac{\psi}{2}\right)$ \\
        \hline
        $\uparrow\downarrow-$&$\sqrt{1-\beta^2}$&$2f_V^Rx^2\beta\sin^2{\psi}$  & $\sqrt{2}x(\kappa_{A+}-2f_V^R \beta\cos{\psi})\sin{\psi}$ &$2(-\kappa_{A+}\cos{\psi}+\kappa_{V+}-f_V^R\beta\sin^2{\psi})$\\
        \hline
        $\downarrow\uparrow-$&$\sqrt{1-\beta^2}$&$2f_V^Rx^2\beta\sin^2{\psi}$  & $\sqrt{2}x(\kappa_{A-}-2f_V^R \beta\cos{\psi})\sin{\psi}$ &$2(-\kappa_{A-}\cos{\psi}-\kappa_{V-}-f_V^R\beta\sin^2{\psi})$\\
        \hline
        $\downarrow\downarrow-$&$4\sin\left(\frac{\psi}{2}\right)$ &$x^2k_L\sin^2{\left(\frac{\psi}{2}\right)}\cos{\left(\frac{\psi}{2}\right)}$ & $-\frac{x}{\sqrt{2}}  \left( \kappa + 2 k_L \cos^2\left(\frac{\psi}{2}\right) \right) \sin\left(\frac{\psi}{2}\right)$ & $- \left( \kappa - k_L \cos^2\left(\frac{\psi}{2}\right) \right)\cos\left(\frac{\psi}{2}\right)$ \\
        \hline
    \end{tabular}
\end{table}

To see the photon radiation tends to happen in the coplanar case, we sum over the helicities and integrate out $\Theta$ (from 0 to $\pi$) and $\psi$ (from $\delta$ to $\pi-\delta$) for the squared amplitude $|\mathcal{M}|^2$ for massless final states, and obtain:
\begin{equation}
\begin{split}
    |\mathcal{M}|^2 &= \frac{e^6}{2 \left(x^2-1\right)^2 c_W^4 s_W^4 \Gamma _Z^2}  \left(64 \left(x^2+1\right) x \log \left(\frac{2}{\delta }\right) \cos (\Phi ) \left(g_L^2-g_R^2\right)^2\right.\\
    &\left.\quad + 4 \pi  \left(g_L^2+g_R^2\right)^2 \left(\frac{6 \left(x^4+1\right)}{\delta }+(\pi -2 \delta ) x^2 \cos (2 \Phi )-\frac{1}{2}(\pi -2 \delta ) \left(3 x^4-4 x^2+3\right)\right).
\right)
\end{split}
        \label{eq:sqrtM}
\end{equation}
We see that the $\Phi$ dependence of the distribution is given by cosine terms and therefore maximizes at $\Phi=0,\pi$, as shown in Fig.~\ref{fig:coplanar_Msqr} with $\delta=0.1$.

\begin{figure}[h]
    \centering
\includegraphics{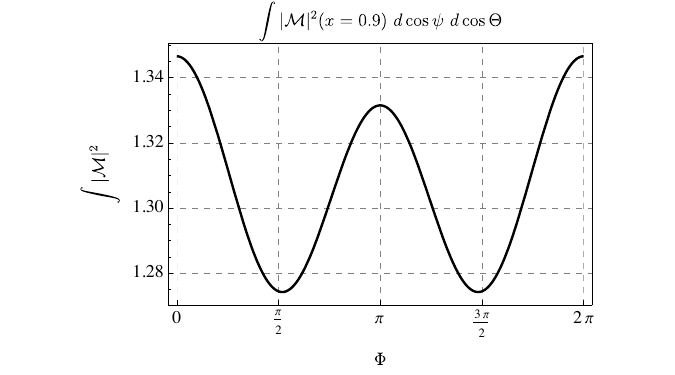}    \caption{$\iint|\mathcal{M}|^2d\cos\Theta~d\cos\psi$ for soft radiation as function of $\Phi$}
 \label{fig:coplanar_Msqr}
\end{figure}

\end{widetext}

\bibliographystyle{apsrev4-1}
\bibliography{ref.bib}

\end{document}